%% file: N1808_AstroPh.tex
\documentclass[rnote]{aa}  
\usepackage{graphicx}
\usepackage{txfonts}
\usepackage{natbib}

\def\kms{$\rm \,km\,s^{-1}$}

\def\HH{H$_2$}

\def\Ha{H${\alpha}$}

\def\Pa{Pa${\alpha}$}

\def\Pg{Pa${\gamma}$}
\def\Pfd{Pf${\delta}$}
\def\Pfg{Pf${\gamma}$}
\def\Ba{Br${\alpha}$}

\def\Bd{Br${\delta}$}
\def\Bg{Br${\gamma}$}

\def\micro{$\rm \,\mu m$}
\def\arcsec{\,''}

\def\flux{$\rm \,erg\,s^{-1}\,cm^{-2}$}

\begin{document}

\title{Embedded clusters in NGC1808 central starburst}

\subtitle{Near-infrared imaging and spectroscopy}

\author{E. Galliano        
        \inst{1,2,3}
        \and	
        D. Alloin 
        \inst{4}
        }


\institute{Observat\'orio Nacional, Rua General Jos\'e Cristino, 77, 20921-400, S\~ao Cristov\~ao, Rio de Janeiro, Brazil\\  
           \email{egallian@on.br}
	   \and 
           Departamento de Astronomia, Universidad de Chile, Casilla 36-D, Santiago, Chile           
           \and
           European Southern Observatory, Casilla 19001, Santiago 19, Chile
	   \and 
	   Laboratoire AIM, CEA/DSM-CNRS-Universit\'e Paris Diderot, IRFU/Service d'Astrophysique, B\^at.709, CEA/Saclay, F-91191 Gif-sur-Yvette Cedex, France
}

\date{Received ---; accepted ---}

\abstract {In the course of a mid-infrared imaging campaign of
  close-by active galaxies, we discovered the mid-infrared
  counterparts of bright compact radio sources in the central
  star-forming region of NGC1808.}  {We aim at confirming that these
  sources are deeply embedded, young star clusters and at deriving
  some of their intrinsic properties.}  {To complement the
  mid-infrared data, we have collected a set of near-infrared data
  with ISAAC at the VLT: J, Ks, and L' images, as well as
  low-resolution, long-slit spectra for three of the sources.}
          {Surprisingly, the new images unveil a near-infrared
            counterpart for only one of the mid-infrared/radio
            sources, namely M8 in the L' band. All the other sources
            are so deeply embedded that their emission does not pop
            out above an extended diffuse near-infrared emission. The
            near-infrared spectra of the sources look alike, with
            intense, ionised hydrogen lines. This supports the
            interpretation of these sources in terms of embedded young
            clusters. We derive extinctions and ionising photon
            production rates for two of the clusters. } {}
           \authorrunning{Galliano \& Alloin}
          \titlerunning{Embedded star clusters in NGC1808} \maketitle

\begin{figure*}[htbp]
\begin{center}
\includegraphics[width=15cm]{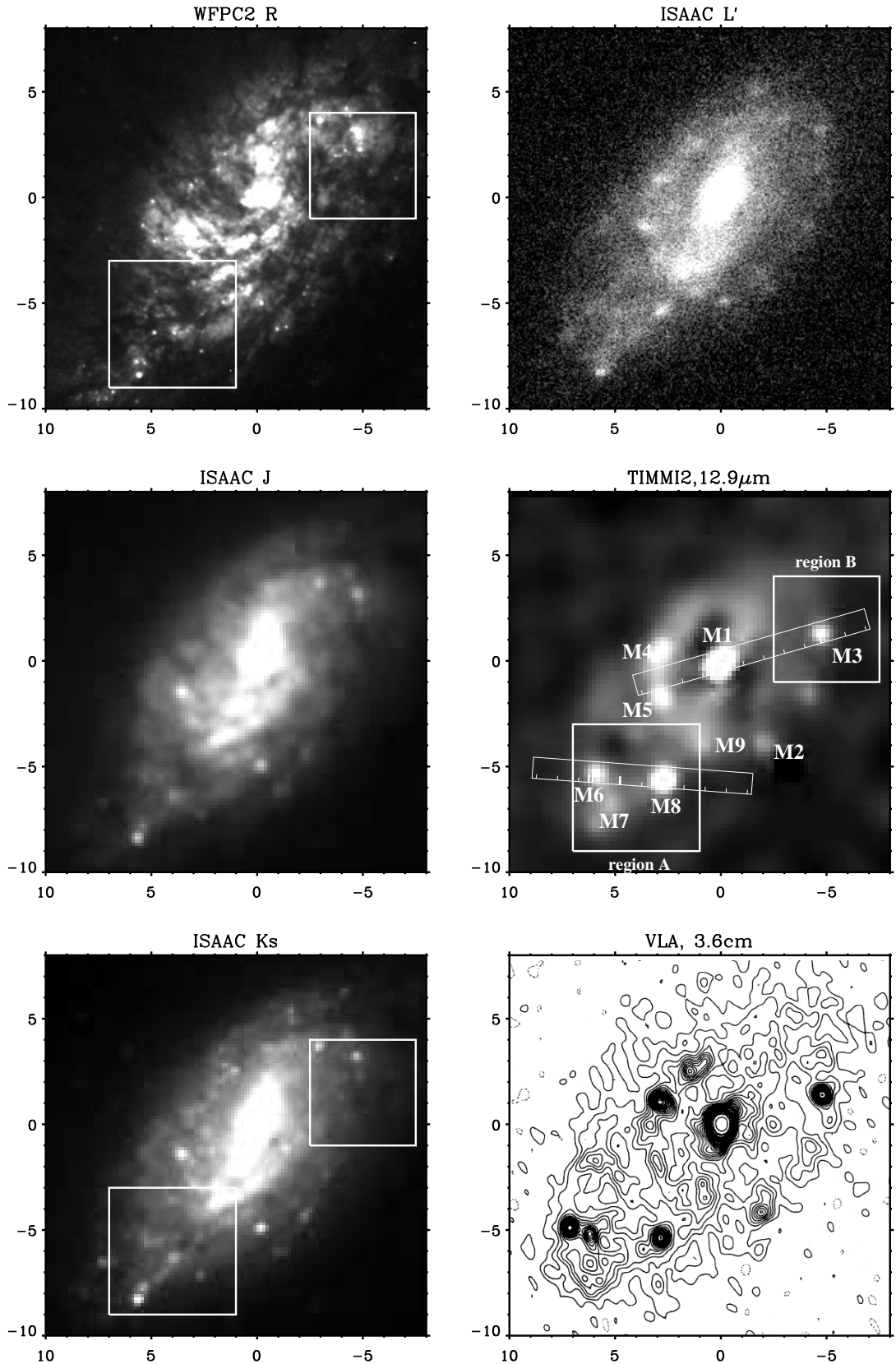}
\caption{The central 18\arcsec$\times$18\arcsec region of NGC1808
  (1\arcsec=53\,pc). The axis units are arcsec. The nucleus (source
  M1) is positioned at coordinates (0\arcsec; 0\arcsec). North is up
  and East is to the left. The MIR sources are labelled as in
  \citet{Galliano05a} on the TIMMI2 map. Zooms towards the two regions
  delimited by the squares (Regions A and B) are displayed in
  Fig.~\ref{fig2}. The long rectangles feature the spectroscopic
  slits: small ticks are spaced by 1\arcsec, while long thick ticks
  highlight the positions at which line emission has been recorded.}
\label{fig1}
\end{center}
\end{figure*}

\begin{figure*}[htbp]
\begin{center}
\includegraphics[width=18cm]{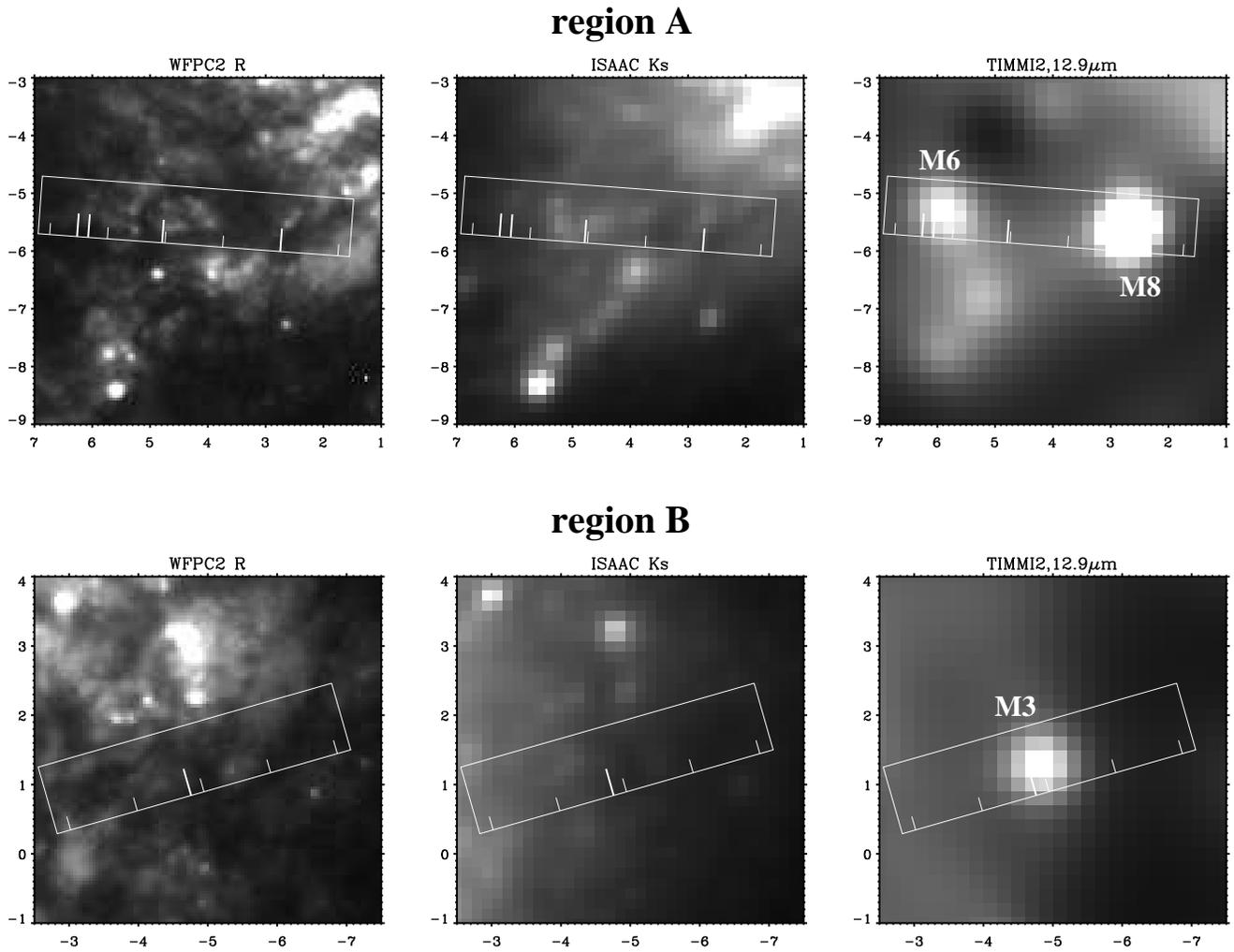}
\caption{Zooms towards regions A and B of the NGC1808 central
  starburst.  These regions are defined in Fig.~\ref{fig1}. The axis
  units are arcsec, North is up and East is to the left. Regarding the
  slit drawings, see the caption of Fig.~\ref{fig1}.}
\label{fig2}
\end{center}
\end{figure*}
 
\begin{figure*}[htbp]
\begin{center}
\includegraphics[width=18cm]{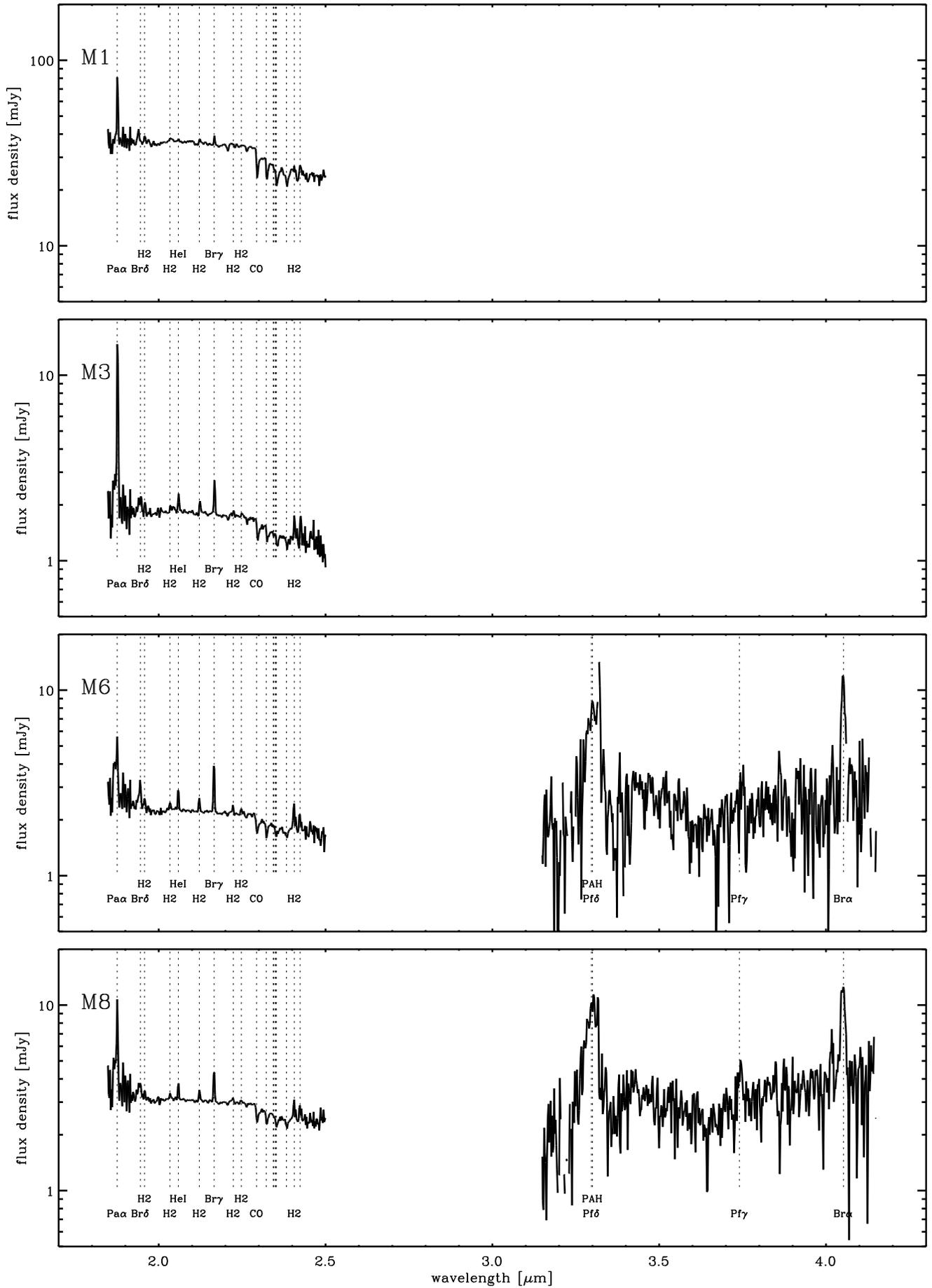}
\caption{ISAAC spectra of the MIR/radio sources M1, M3, M6 and M8 in the
  central region of NGC1808. The slit positions, as well as the source
  positions along the slits are shown in Fig.~\ref{fig1}
  and~\ref{fig2}.}
\label{fig3}
\end{center}
\end{figure*}

\begin{figure}
\begin{center}
\includegraphics[width=8cm]{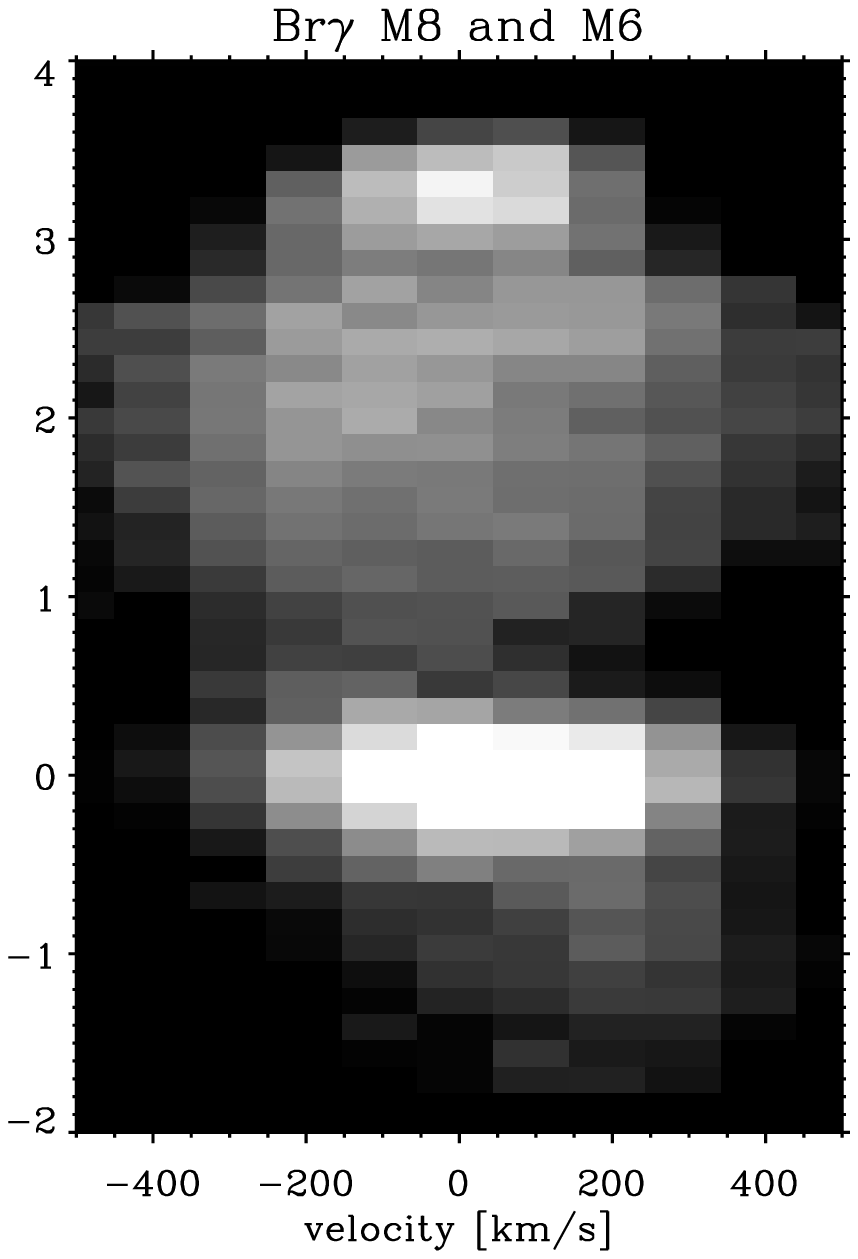}
\caption{Position-velocity diagram for Br${\gamma}$ along the slit
  which contains the two embedded clusters M6 and M8. The X-axis shows
  the wavelength expressed in relative radial velocity in \kms~and the
  Y-axis gives the relative angular position in arcsec.}
\label{fig4}
\end{center}
\end{figure} 

\input{table_N1808_fin3}

\section{Introduction}
\label{Introduction}
Star clusters appear to be the most frequent mode of star formation in
galaxies. Therefore, the understanding of star cluster formation and
evolution is necessary for studying the stellar populations in
galaxies in general. Young clusters in starburst environments have
been extensively studied since the beginning of the nineties, thanks
to the high angular resolution of the Hubble space telescope
(HST). Yet, the very early stages of their evolution are still poorly
known, as they are expected to form in dense dusty environments, hence
to suffer high extinction. As it is only recently that high angular
resolution instruments in the near- and mid-infrared (NIR, MIR) have
become available, only a small sample of embedded young clusters is
known.
 
In a previous study \citep{Galliano05a} we discovered bright MIR
sources within the central starburst of NGC1808, associated with
already known radio sources. We suggested that these sources were
young embedded star clusters. The present research note aims at
complementing the MIR dataset with new NIR data for these sources. We
present NIR images of the central starburst, as well as NIR long slit
spectra of three of the detected MIR/radio sources. The spectra show
intense nebular emission lines and therefore bring strong support to
an interpretation of the sources in terms of young embedded
clusters. From the spectral measurements, some basic parameters such
as extinction and ionising photon emission rate are derived
for the embedded clusters.

NGC1808 is located at a distance of 10.9\,Mpc \citep{Koribalski96},
which corresponds to a scale of 53\,pc per arcsec. Its redshift is
0.003319 \citep{Koribalski04}. The central region (inner 750\,pc) of
this galaxy is undergoing an intense episode of star formation,
discussed in e.g.~\citet{Tacconi-Garman96,Tacconi-Garman05}.

\section{Observations}
\label{Observations}

With ISAAC at VLT/UT1, we collected NIR images and spectra of the
circumnuclear MIR sources discovered in NGC1808 \citep{Galliano05a}.
We adopt hereafter the source nomenclature as in \citet{Galliano05a}.
We recorded three images: J (1.2\micro)\footnote{P074.B-0166; October
  29-30, 2004}, Ks (2.2\micro)~and L'
(3.8\micro)\footnote{P072.B-0397; December 01-02, 2003}. The image
angular resolution is around 0.6\arcsec. We also obtained low
resolution long slit spectra for two slit positions. For one slit
position (passing through sources M6 and M8), we obtained a
2\micro~spectrum (1.8-2.4\micro) and a 4\micro~spectrum
(3.1-4.1\micro).  For the other slit position (across the nucleus M1
and source M3), we could only obtain a 2\micro~spectrum. The spectral
resolutions are R=450 and 360 for the 2\micro~and 4\micro~spectra
respectively. The slits were positioned performing offsets referenced
on the nucleus M1, according to the measurements in
\citet{Galliano05a}. The data reduction was performed using the
\texttt{ECLIPSE} dedicated routines and \texttt{IRAF}. The spectra
were extracted through slit windows of 1.4\arcsec~along the slit.  In
addition to these data, we use in this research note an archive,
pipe-line reduced, WFPC2 narrow filter F658N (655\,nm-661\,nm) image
of the starburst, as well as the TIMMI\,2 10.4, 11.9 and
  12.9\micro~images~presented in \citet{Galliano05a}.

\section{Results}
\subsection{Images}
\label{Images}
Figure~\ref{fig1} shows the central starburst in NGC1808
(18\arcsec$\times$18\arcsec, corresponding to 1\,kpc$\times$1\,kpc) at
six different wavelengths from visible to radio: 0.66\micro,
1.2\micro, 2.2\micro, 3.8\micro, 12.9\micro~and 3.6\,cm. The bulk of
the emission consists of an 20\arcsec$\times$10\arcsec~emitting region
roughly aligned along PA=135\degr. In the maps shown in
Fig.~\ref{fig1}, the coordinates (0\arcsec; 0\arcsec) correspond to
the brightest source M1. The nature of this source has not yet
  been clarified in the literature. It is suggested to be a mixture
  of star formation and a low luminosity or fading active galactic
  nucleus \citep{Krabbe01,Jimenez-Bailon05}. This source is used for
the relative registration of the different maps, assuming that the
luminosity peak coincides at all studied wavelengths.  The following
images are displayed in Fig.~\ref{fig1}, in order of increasing
wavelength:

\begin{itemize}
\item the narrow filter WFPC2 images (0.66\micro), which, at the
  redshift of NGC1808 traces the \Ha~emission, and is hence a
  good tracer of un-embedded star formation in the region,
\item the J (1.2\micro) and Ks (2.2\micro) images which trace the
  photospheric emission of stars, less subject to extinction than in
  the visible,
\item the L' band image (3.8\micro) which can highlight hot dust, and
  hence globally traces star forming activity and AGN activity,
\item the 12.9\micro~image which traces warm dust emission and [NeII]
  line emission.  As for the L' band emission, the image pinpoints
  star forming regions and the AGN,
\item the 3.6\,cm image which corresponds to the free-free emission
  from HII regions, to the diffuse synchrotron emission produced by
  electrons escaping SN remnants and being acceleration by galactic
  magnetic fields, and to the likely emission from recent SN
  explosions.
\end{itemize}

The maps on Fig.~\ref{fig1} bring new details about the extended star
forming region. Through its \Ha~emission (WFPC2 image), the region
looks extremely patchy, likely the result of a complex distribution of
star formation ``nests'' and of molecular/dusty clouds producing a
heavy extinction. It is interesting to note the spiral arm-like
structure of the F658N emission, particularly to the South of the
nucleus. The J and Ks images do not add much information to this
picture, except for an enhanced contrast for some sources which look
faint inn the R image. They do not unveil any new source, with respect
to the R map.

In the L' band, the region exhibits a bright diffuse emission,
possibly the signature of a large amount of dust heated by the various
star forming nests. Several compact sources appear on this map.

The following two images (12.9\micro~and 3.6\,cm) show deeply embedded
sources which do not appear on either the R, J, Ks or L' maps. This
remark is important, showing that embedded star forming activity in
starbursts can be hidden even at 4\micro. As already pointed out in
\citet{Galliano05a}, the correlation between the 12.9\micro~map and
the 3.6\,cm map is excellent.

Figure~\ref{fig2} offers zooms on two regions of interest (regions A
and B) both containing MIR/radio sources. In region A, M8 is detected
in L' only, while M6 is not even detected in L'. None of these sources
show up in the NIR. Regarding region B, there is no detection either
in Ks or L'. Strikingly, on the Ks map, the location of M3 even
corresponds to a dark area in the diffuse emission.

Conversely, we note that the K knots detected by
\citet{Tacconi-Garman96}, and which are interpreted as being embedded
young stellar clusters too, are not detected on the 12.9\micro~map.

Hence, there are two types of embedded clusters in NGC1808: one bright
in the MIR, bright in the radio and faint in the NIR, and the other
bright in the NIR and faint in the MIR and the radio. This dichotomy
can simply be explained if we consider that MIR/radio clusters are at
an earlier evolution stage than the sole K clusters: at a few million
years of age, the clusters are still deeply embedded in their
formation material, which is a mixture of ionised gas (source of the
radio emission and nebular emission lines) and dust (source of the MIR
continuum). At this stage, the stars are so heavily embedded that they
do not appear even on the K maps. This locally embedding material is
eventually expelled, but the clusters still reside within the kpc
scale dusty star formation region, hence suffer some extinction. At
this stage, the clusters appear as K extincted sources, with no MIR or
radio counter parts. This interpretation is further supported by
\citet{Tacconi-Garman05}: not only is the PAH-to-continuum ratio at
the K knot positions low, but also it is high at the radio/MIR knot
positions \citep[See Fig.~4b~in~][]{Tacconi-Garman05}. This could mean
that much of the PAH/dust around the K knots is photoionised, hence
weakening the 3.3\micro~PAH feature, has been photodissociated or has
simply been blown away, while this has not yet happened in the
MIR/radio knots.

\subsection{Spectra}
\label{Spectra}
We have obtained NIR spectra of three embedded clusters and the
nucleus: 2\micro~and 4\micro~spectra for M6 and M8, and only one
2\micro~spectrum for M1 and M3. The projected slit positions are drawn
in the images in Fig.~\ref{fig1} and Fig.~\ref{fig2}. On these
drawings, small ticks mark each arcsec, while the long and thick ticks
mark the positions at which line emission has been recorded. We detect
emission lines for M1 (the nucleus), M3, M6 and M8, and we even detect
line emission in the region between M6 and M8 devoid of any
conspicuous MIR or radio source. The spectra obtained for M1, M3, M6
and M8 are shown in Fig.~\ref{fig3}.

For M1, the continuum shown on Fig.~\ref{fig3} is, at the achieved
angular resolution, associated to the emission line source. Three
emission lines are un-ambiguously detected and measurable: \Pa, H2
1-0S(1) and \Bg. The HeI 2.058\micro~is only tentatively detected and
its flux cannot be measured because of the presence of artificial
features on this part of the spectrum. Deep 2.3\micro~CO absorption
bands are detected, tracing the presence of red super giants. The
equivalent width of \Bg~is 4.5\AA. This can be compared to Starburst99
predictions \citep{Leitherer99}, but we must keep in mind that part of
the continuum, even in K, may arise from hot dust associated to the
AGN. Hence, the measured value of the equivalent width of \Bg~is a
lower limit. By comparing with Fig. 89 and 90 of \citet{Leitherer99},
we deduce that (1) in the instantaneous star formation case, such an
equivalent width implies that the cluster is younger than 7\,Myr; (2)
in the continuous star formation case, there is no age constraint,
since we ignore the AGN 2\micro~continuum share. In conclusion, the K
M1 spectrum does not clarify the nature of the source.

In the following, we focus on the three circumnuclear sources M3, M6
and M8.  Their spectra look quite similar in shape and are typical HII
region spectra. Emission lines from atomic (\Pa, \Bd~and \Bg) and
molecular hydrogen show up. In M6 and M8, for which we have an
additional L' spectrum, we detect the hydrogen lines \Pg~and \Ba~as
well as the PAH band at 3.3\micro.  This strongly supports an
interpretation of the MIR/radio sources in terms of heavily embedded
clusters as already suggested in \citet{Galliano05a}. In addition,
such spectra fix an upper limit of 10\,Myrs to the cluster ages,
otherwise no nebular gas emission would be seen \citep{Leitherer99}.
The continuum shown on the spectra of M3, M6 and M8 is not associated
with the sources of emission lines. This continuum only shows the
diffuse emission of the starbursting region. While the continuum
emission is extended, the emission lines are only observed at the
location of the MIR sources along the slit. Hence, the CO absorption
bands seen on the spectrum continua are not associated to the embedded
clusters.

Fig.~\ref{fig4} displays a 2-D spectrum for the \Bg~line along the
M6-M8 direction. The line emission extends over several arcsec
($\sim$~250\,pc) and exhibits a complex velocity structure, suggestive
of an outflow of ionised material. This probably witnesses the
  expelling of the locally embedding material suggested in
  Sect.~\ref{Images}.

\subsection{Measurements}
\label{Measurements}
Table~\ref{table measurements} provides a summary of the measurements
performed on the data and gives the radio measurements of
\citet{Collison94}. The following procedure has been followed for the
flux density measurements and the uncertainty estimates. The relative
positions of the three sources with respect to M1 are taken as those
in the TIMMI2 12.9\micro~image. They are given in the table. At these
positions, we measure the flux densities inside a 0.6\arcsec~radius
aperture. The background is estimated by computing the median within
an annulus with radii 0.6\arcsec~and 1.2\arcsec. Since, for these
data, the background is difficult to interpret, we consider two values
of the flux density for each source, one with and one without
background subtraction, this is what we call the measurement error. We
then compute an error defined as the quadratic sum of the measurement
error and the photometric calibration uncertainty. The following flux
calibration uncertainties are considered: 10$\%$ for the R, J and Ks
images and 20$\%$ for the L' and MIR images. The two values presented
in table~\ref{table measurements} correspond to the final lower and
upper limits of each measurement. When no clear identification of the
source in the considered aperture is possible (for example, for the
data in the R, J and Ks band), we did not consider the lower limit
defined above, but only the upper limit. For the line fluxes, we
consider a 20\% uncertainty, dominated by the
calibration. Nevertheless, for the lines on the blue side of the
2\micro~spectrum (\Pa, \Bd~and H2 1-0S(3)), the uncertainty may be
greater than this value because of the strong atmospheric features
present in this wavelength range and the values given must be taken
with caution.

\section{Cluster derived parameters}
\label{Derived parameters}

In this section, we discuss some intrinsic parameters of the sources
which can be derived from the NIR emission lines. The procedure has
been applied on M6 and M8, for which both the 2\micro~spectra
(\Bg~flux) and 4\micro~spectra (\Ba~flux) are available.  Under the
assumption of simple foreground extinction, we derive the extinction
towards the line emission gas, and then compute the un-extincted line
luminosities.  This in turn allows an estimation of the ionising photon
emission rate.

We have deduced the extinction from the \Bg/\Ba~ratio. This ratio has
a low dependency on temperature and density. Assuming reasonable
values for the temperature and density of HII regions, $\rm
T_e=10^{4}$\,K and $\rm N_e=10^{4}\,cm^{-3}$ \citep{Osterbrock89}, we
obtain figures of Av=10.2 and 15.0, for M6 and M8 respectively.  This
leads to un-extincted \Bg~fluxes of 8.9$\times$10$^{-15}$\flux~ in M6,
and 9.5$\times$10$^{-15}$\flux~ in M8.

For an embedded young star cluster, the ionisation-bounded case is
very likely: the HII region is optically thick in the Lyman continuum
and all the stellar ionising photons are absorbed. Then, the number of
hydrogen ionising photons ($\lambda \le 912\AA$) is directly
proportional to the flux in any specific recombination line.

From \citet{Osterbrock89},
\begin{equation}
\rm Q[H^+]=\frac{\alpha_B}{\alpha^{eff}_{H\alpha}} \times
\frac{L_{H\alpha}}{h\nu_{H\alpha}}~~,~where~\alpha_B/\alpha^{eff}_{H\alpha}\sim
2.96.
\end{equation}

{\noindent In case B, we get $\rm L_{H\alpha}/L_{B\gamma}$= 103.6. The ionising
photon emission rates of the two embedded clusters can then be derived
and are found to be around 9.6$\times 10^{51}$\,s$^{-1}$ and
10.2$\times 10^{51}$\,s$^{-1}$, in M6 and M8 respectively. This
corresponds to about 1000 O6 stars in each cluster.}

The way we have treated extinction in the above computation is a crude
one. This is undoubtly a first order estimate which calls for a
future, more realistic approach, once additional data constraints will
become available. Yet, the interpretation of the MIR/radio sources in
terms of embedded young clusters is confirmed and, in addition, an
upper limit on the cluster ages is obtained at 10\,Myrs.

\section{Conclusions}
The main conclusions are:
\begin{itemize}
\item None of the MIR/radio sources unveiled by \citet{Galliano05a}
  can be detected in the J, Ks and even L' band images, except for M8
  in the L' band.
\item For three of the sources, namely M3, M6 and M8, we have collected NIR 
spectra: they display intense nebular emission lines, confirming that the 
sources are young embedded clusters. 
\item The fact that the MIR/radio sources do not show up in the NIR images 
demonstrates that such embedded clusters must be searched for primarily in 
the MIR or at radio wavelengths. 
\item A complex velocity structure in the ionised gas is observed around 
sources M6 and M8, suggesting the presence of cluster winds. 
\item For two sources, M6 and M8, the extinction towards the nebular gas 
has been estimated, with values of Av=10.2 in M6 and Av=15 in M8. 
\item The extinction-corrected \Bg~fluxes for M6 and M8 imply ionising photon
  emission rates of respectively 9.6\,$\times$$10^{51}$\,s$^{-1}$ and
  10.2\,$\times$$10^{51}$\,s$^{-1}$. 
\end{itemize}

A more complete analysis of the deeply embedded cluster parameters, in
particular the derivation of their age and mass, requires some
knowledge of their MIR spectral features \citep{Galliano07a} and it is
left for a subsequent analysis, after spectral MIR data will have been
collected.

\begin{acknowledgements}
We thank the daytime and nighttime support staff at Cerro Paranal
Observatory, who made these observations possible, and the anonymous
referee for the quick reply. EG thanks the ESO fellowship program and
the PCI program of ON/MCT (DTI/CNPq grant number 383076/07-2). 
\end{acknowledgements}

\bibliographystyle{aa}

\end{document}

%% file: table_N1808_fin3.tex
\begin{table*}[htbp]
\begin{center} 
\caption[]{Measurements for the clusters M1, M3, M6 and M8 in NCG1808}
\label{table measurements}
\begin{tabular}{lcccccccc}

 & \multicolumn{2}{c}{M1}   & \multicolumn{2}{c}{M3}   & \multicolumn{2}{c}{M6}   & \multicolumn{2}{c}{M8}\\ 
\vspace{0.01cm}\\
\hline

\vspace{0.01cm}\\
\multicolumn{9}{c}{positions with respect to M1 [arcsec]}\\
\vspace{0.01cm}\\
$\Delta\alpha$~[arcsec]& \multicolumn{2}{c}{0} & \multicolumn{2}{c}{-1.99$\pm$0.1} & \multicolumn{2}{c}{5.82$\pm$0.1} & \multicolumn{2}{c}{2.71$\pm$0.1} \\ 
$\Delta\delta$~[arcsec]& \multicolumn{2}{c}{0} & \multicolumn{2}{c}{-3.76$\pm$0.1} & \multicolumn{2}{c}{-5.25$\pm$0.1} & \multicolumn{2}{c}{-5.51$\pm$0.1} \\ 
\vspace{0.01cm}\\
\hline

\vspace{0.01cm}\\
\multicolumn{9}{c}{flux densities [mJy]}\\
 & \multicolumn{2}{c}{value}& \multicolumn{2}{c}{value}& \multicolumn{2}{c}{value}& \multicolumn{2}{c}{value}\\ 
             & low    & high    & low        & high     & low      & high     & low      & high     \\ 
\vspace{0.01cm}\\
R            &    2.2 &     3.1 &     0      &     0.2  &     0    &     0.2  &     0    &     0.2  \\                                                                                 
J            &   11.  &    20.  &     0      &     1.2  &     0    &     1.3  &     0    &     1.7  \\                                                                                 
Ks           &   22.  &    39.  &     0      &     2.0  &     0    &     2.2  &     0    &     2.9  \\                                                                                 
L'           &   17.  &    38.  &     0      &     2.8  &     0    &     3.4  &     0.4  &     4.8  \\                                                                                 
10.4\micro~  &  180.  &   280.  &     0      &    12.   &     0    &    11.   &     6.4  &    23.   \\                                                                                 
11.9\micro~  &  400.  &   670.  &    11.     &    50.   &    10.   &    76.   &    18.   &    99.   \\                                                                                 
12.9\micro~  &  700.  &  1100.  &    25.     &    90.   &    29.   &   110.   &    95.   &   220.   \\                                                                                 
\vspace{0.01cm}\\
\hline

\vspace{0.01cm}\\
\multicolumn{9}{c}{line fluxes [$\rm 10^{-15}\,erg\,s^{-1}\,cm^{-2}$]} \\
\vspace{0.01cm}\\
$^*$\Pa      &\multicolumn{2}{c}{  150.$\pm$   30.    } & \multicolumn{2}{c}{   49.$\pm$    10.   } & \multicolumn{2}{c}{    9.9$\pm$    2.0   } & \multicolumn{2}{c}{   24.$\pm$     5.} \\ 
$^*$\Bd      &\multicolumn{2}{c}{  $<$4.              } & \multicolumn{2}{c}{   $<$1.4            } & \multicolumn{2}{c}{    3.9$\pm$    0.8   } & \multicolumn{2}{c}{    2.8$\pm$    0.6} \\ 
\HH~1-0S(3)  &\multicolumn{2}{c}{    $<$12.           } & \multicolumn{2}{c}{   $<$1.2            } & \multicolumn{2}{c}{    $<$1.6            } & \multicolumn{2}{c}{    $<$1.6         } \\ 
\HH~1-0S(2)  &\multicolumn{2}{c}{    $<$10.           } & \multicolumn{2}{c}{   $<$0.6            } & \multicolumn{2}{c}{    0.8$\pm$    0.2   } & \multicolumn{2}{c}{    1.1$\pm$    0.3} \\ 
HeI          &\multicolumn{2}{c}{   $<$5.             } & \multicolumn{2}{c}{    1.3$\pm$    0.3  } & \multicolumn{2}{c}{    2.1$\pm$    0.5   } & \multicolumn{2}{c}{    2.2$\pm$    0.5} \\ 
\HH~1-0S(1)  &\multicolumn{2}{c}{    5.8$\pm$    1.2  } & \multicolumn{2}{c}{    0.9$\pm$    0.2  } & \multicolumn{2}{c}{    1.2$\pm$    0.3   } & \multicolumn{2}{c}{    1.3$\pm$    0.3} \\ 
\Bg          &\multicolumn{2}{c}{    9.6$\pm$    1.9  } & \multicolumn{2}{c}{    2.6$\pm$    0.6  } & \multicolumn{2}{c}{    5.4$\pm$    1.1   } & \multicolumn{2}{c}{    4.2$\pm$    0.9} \\ 
\HH~1-0S(0)  &\multicolumn{2}{c}{   $<$2.             } & \multicolumn{2}{c}{   $<$0.5            } & \multicolumn{2}{c}{    0.6$\pm$    0.2   } & \multicolumn{2}{c}{    0.8$\pm$    0.2} \\ 
\HH~2-1S(1)  &\multicolumn{2}{c}{   $<$2.             } & \multicolumn{2}{c}{   $<$0.5            } & \multicolumn{2}{c}{    0.6$\pm$    0.2   } & \multicolumn{2}{c}{   $<$6.5          } \\ 
\HH~1-0Q(1)  &\multicolumn{2}{c}{   $<$5.5            } & \multicolumn{2}{c}{   1.0$\pm$0.2       } & \multicolumn{2}{c}{    1.7$\pm$    0.4   } & \multicolumn{2}{c}{    1.5$\pm$    0.3} \\ 
\HH~1-0Q(3)  &\multicolumn{2}{c}{   $<$5.5            } & \multicolumn{2}{c}{   1.2$\pm$0.3       } & \multicolumn{2}{c}{    1.0$\pm$    0.2   } & \multicolumn{2}{c}{    1.1$\pm$    0.3} \\ 
PAH+\Pfd     &\multicolumn{2}{c}{   --                } & \multicolumn{2}{c}{   --                } & \multicolumn{2}{c}{   84.$\pm$    17.    } & \multicolumn{2}{c}{  100.$\pm$    20. } \\ 
\Pfg         &\multicolumn{2}{c}{   --                } & \multicolumn{2}{c}{   --                } & \multicolumn{2}{c}{   3.2  $\pm$    0.7  } & \multicolumn{2}{c}{   5.4  $\pm$1.1   } \\ 
\Ba          &\multicolumn{2}{c}{   --                } & \multicolumn{2}{c}{   --                } & \multicolumn{2}{c}{   20.$\pm$    4.     } & \multicolumn{2}{c}{   25.$\pm$     5. } \\ 

\vspace{0.01cm}\\
\hline

\vspace{0.01cm}\\
\multicolumn{9}{c}{radio data \citep{Collison94}}\\
\vspace{0.01cm}\\
$\alpha^{\rm 6\,cm}_{\rm 3.6\,cm}$ & \multicolumn{2}{c}{-0.61} & \multicolumn{2}{c}{-0.28} & \multicolumn{2}{c}{--} & \multicolumn{2}{c}{-0.57} \\
3.6\,cm [mJy] & \multicolumn{2}{c}{8.96} & \multicolumn{2}{c}{0.93} & \multicolumn{2}{c}{0.58} & \multicolumn{2}{c}{0.76} \\
\end{tabular}
\end{center}
$^*$ These flux values and their associated uncertainties must be taken with caution since the atmospheric features are strong in this region of the spectrum. 
\end{table*}